# PHASE DIAGRAMS AND DOMAIN SPLITTING IN THIN FERROELECTRIC FILMS WITH INCOMMENSURATE PHASES

A.N. Morozovska[*,1], E.A. Eliseev[1,2], JianJun Wang[3], G.S. Svechnikov[1], Yu. M. Vysochanskii[4], Venkatraman Gopalan[3], and Long-Qing Chen[3,†]

[1]V. Lashkarev Institute of Semiconductor Physics, NAS of Ukraine,
prospect Nauki 41, 03028 Kiev, Ukraine

[2]Institute for Problems of Materials Science, NAS of Ukraine,
Krjijanovskogo 3, 03142 Kiev, Ukraine

[3]Department of Materials Science and Engineering, Pennsylvania State University,
University Park, Pennsylvania 16802, USA

[4]Institute for Solid State Physics and Chemistry, Uzhgorod University,
88000, Uzhgorod, Ukraine

**Abstract**

We studied the phase diagram of thin ferroelectric films with incommensurate phases and semiconductor properties within the framework of Landau-Ginzburg-Devonshire theory. We performed both analytical calculations and phase-field modelling of the temperature and thickness dependencies of the period of incommensurate 180º-domain structures appeared in thin films covered with perfect electrodes. It is found that the transition temperature from the paraelectric into the incommensurate phase as well as the period of incommensurate domain structure strongly depend on film thickness, and surface and gradient energy contributions. The results may provide insight on the temperature dependence of domain structures in nanosized ferroics with inherent incommensurate phases.

## I. Introduction

The influence of surfaces and interfaces on ferroic materials and their domain structure have been attracting much attention since early seventies till the present.[1, 2, 3, 4, 5, 6, 7] Laminar domain structure formation in thick films with free surfaces was considered in the classic papers by Kittel[8] for ferromagnetic media and Mitsui and Furuichi[9] for ferroelectric media. The structure of a single

---

[*] corresponding author: morozo@i.com.ua
[†] corresponding author: lqc3@psu.edu

boundary between two domains in the bulk ferroelectrics was considered by Cao and Cross[10] and Zhirnov[11], allowing for electrostriction contribution. Formation and stability of ferroelastic domain structures were considered by different groups.[12, 13, 14]

The development of non-volatile ferroelectric memory technology has rekindled the interest in ferroelectric properties and polarization reversal mechanisms in ultrathin films.[15, 16, 17, 18] One of the key parameters controlling ferroic behavior is the structure and energetic of domain walls.

The wall behavior at surfaces and interfaces will determine polarization switching and pinning mechanisms. Under the absence of external fields in the bulk, the $180^o$-domain wall is not associated with any depolarization effects. However, the symmetry breaking on the wall-surface or wall-interface junction can give rise to a variety of unusual effects due to the depolarization fields across the wall, as determined by screening mechanisms and strain boundary conditions.[19] For instance, recent Density Functional Theory simulation results predicted the stabilization of vortex structure in ferroelectric nanodots under the transverse inhomogeneous static electric field.[20, 21] This prediction has resulted in extensive experimental efforts to discover toroidal polarization states in ferroelectrics.[22, 23]

However, despite numerous studies, size and surface effect on domain walls behavior in ferroics is still not clear. Much remained to be done to clarify the peculiarities of the order parameter redistribution in the wall vicinity, corresponding wall energy and walls interaction energy in confined systems like thin films and nanoparticles. For instance, simple analytical models typically face with "Kittel paradox": 180-degree "rigid" domain structure with ultra-sharp walls produces extra-high depolarization fields near unscreened surface[24]. Possible formation of closure domains in "rigid" ferroelectrics with infinitely thin domain walls does not solve the problem. Relevant analytical treatment of multi-axial polarization switching allowing for domain walls intrinsic widths is still underway due to the numerous obstacles. At the same time both first principle calculations and phenomenological modeling revealed unusual domain structures in different ferroelectrics[25, 26], resembling domain structures typical for ferromagnetics.

The incommensurate phase in bulk materials is the spatially modulated state with period incommensurate with the lattice constant.[27] The spontaneous modulation appears when the homogeneous state is either unstable or less energetically preferable (metastable). On the other hand, the initial homogenous states could become modulated in the spatially confined systems. Typical examples are domain structures of either ferroelectric or ferromagnetic films due to the depolarization or demagnetization fields respectively.

The evolution of domain structure in thin films and nanoparticles with incommensurate phase (corresponding to negative gradient energy resulting into effective negative domain wall energy) actually was not considered theoretically. Possibly it is due to the fact the situation with the theoretical description of the incommensurate ferroelectrics is much more complex in comparison with the commensurate ones. In particular within Landau phenomenological approach of the II-type incommensurate materials, the characteristics of modulated phase should be found from forth order nonlinear Euler-Lagrange differential equations (see e.g. Refs. [28, 29, 30]); for commensurate ferroelectrics the equations are of the second order. The most intriguing feature is the mechanisms of commensurate-incommensurate phase transitions. The transition in three-dimensional solids was considered as lock-in transition from the incommensurate phase with negative energy of domain walls into the commensurate phase with positive energy of domain walls[31]. Levanyuk et al. pointed out that electrostriction coupling between polarization and strains significantly changes the phase equilibrium.

The link between the phenomenological model of incommensurate crystals and quasi-microscopic discrete lattice model was established in Ref.[32]. The temperature dependence of the polarization wave number in ferroelectric $Sn_2P_2Se_6$ as well as the anomalous heat capacity in the incommensurate phase were explained in the framework of II-type phenomenological theory using the non-harmonic distribution of the order parameter.[30] First principle calculations[33] may pour light of the local structure of the incommensurate ferroelectrics, however their realization for confined systems like thin films are almost not evolved to date.

It should be expected that size effects essentially influence on the commensurate-incommensurate phase transitions in nanoparticles and thin films. However, the phenomenological theory of the phase transitions in confined incommensurate systems was not evolved. Only few papers were published. Namely, Charnaya et al. [34] obtained the order parameter distribution over the film using the assumption of slowly-varying amplitude and considered the size effect on the temperature of the phase transition into the incommensurate phase. However, the direction of incommensurate phase modulation was normal to the film plane. Since the depolarization field was not considered, this actually means that a polarization vector has only in-plane component, while the films with out of plane polarization is out of the model. In the latter case the problems becomes essentially two-dimensional and the depolarization field becomes inevitable present in the system. Only for the case of an ideal screening and the absence of surface energy dependence on polarization (i.e. for so-called natural boundary conditions) the depolarization field is absent and bulk modulated phase is recovered for thin films.

The paper is organized as follows. Section II is the problem statement. Here we listed expressions for the depolarization field and the free energy functional of ferroelectric thin films with II-type incommensurate phase and semiconductor properties. Approximate analytical solution of the Euler-Lagrange equations is presented in Section III. Results of the analytical calculations of the size effect on phase equilibrium and domain structure temperature evolution are discussed in Section IV.1. Section IV.2 contains results obtained by phase-field modeling. Last section is a brief summary. Mathematical details are summarized in Appendixes.

## II. Phenomenological description of the ferroelectric thin films with the II-type incommensurate phase

Let us consider an incommensurate ferroelectric film with inhomogeneous spontaneous polarization and semiconductor properties. The spontaneous polarization $P_3$ is directed along the polar axis z. The sample is dielectrically isotropic in transverse directions, i.e. permittivity $\varepsilon_{11} = \varepsilon_{22}$ at zero external field.

Further we assume that the dependence of in-plane polarization components on $E_{1,2}$ can be linearized as $P_{1,2} \approx \varepsilon_0(\varepsilon_{11} - 1)E_{1,2}$ ($\varepsilon_0$ is the universal dielectric constant). Thus the polarization vector acquires the form: $\mathbf{P}(\mathbf{r}) = (\varepsilon_0(\varepsilon_{11} - 1)E_1, \varepsilon_0(\varepsilon_{11} - 1)E_2, P_3(\mathbf{E},\mathbf{r}) + \varepsilon_0(\varepsilon_{33} - 1)E_3)$.[35] Maxwell's equations for the inner electric field $\mathbf{E}_i = -\nabla \varphi_i(\mathbf{r})$, expressed via electrostatic potential $\varphi_i(\mathbf{r})$ and polarization $\mathbf{P}(\mathbf{r})$ reduces to the equation $\dfrac{\partial^2 \varphi_i}{\partial z^2} + \dfrac{\varepsilon_{11}}{\varepsilon_{33}}\left(\dfrac{\partial^2 \varphi_i}{\partial x^2} + \dfrac{\partial^2 \varphi_i}{\partial y^2}\right) - \dfrac{\varphi_i}{\varepsilon_{33} R_d^2} = \dfrac{1}{\varepsilon_0 \varepsilon_{33}} \dfrac{\partial P_3}{\partial z}$, $R_d$ is the Debye screening radius. A background permittivity $\varepsilon_{33}$ is regarded much smaller than ferroelectric contribution to temperature-dependent permittivity $\varepsilon_{33}^f$.

The boundary conditions $\varphi_i(x,y,0) = \varphi_e(x,y,0)$, $\varphi_i(x,y,h) = 0$ used hereinafter correspond to the full screening of depolarization field outside the sample that is realized by the ambient charges or perfect electrodes; $h$ is the film thickness.

In Debye approximation the Fourier representation on transverse coordinates $\{x,y\}$ for the depolarization field $E_3^d$ has the form (see Appendix A for details):

$$\tilde{E}_3^d\left[\tilde{P}_3(\mathbf{k},z)\right] = \begin{pmatrix} -\dfrac{\tilde{P}_3(\mathbf{k},z)}{\varepsilon_0 \varepsilon_{33}} + \int_0^z dz' \tilde{P}_3(\mathbf{k},z')\cosh(K z')\dfrac{\cosh(K(h-z))}{\varepsilon_0 \varepsilon_{33} \cdot \sinh(K h)} K + \\ \int_z^h dz' \tilde{P}_3(\mathbf{k},z')\cosh(K(h-z'))\dfrac{\cosh(K z)K(k)}{\varepsilon_0 \varepsilon_{33} \cdot \sinh(K h)} \end{pmatrix} \quad (1)$$

Here vector $\mathbf{k} = \{k_1, k_2\}$, its absolute value $k = \sqrt{k_1^2 + k_2^2}$, function $K(k) = \sqrt{(\varepsilon_{11} k^2 + R_d^{-2})/\varepsilon_{33}}$. For the transversally homogeneous dielectric media with $R_d \to \infty$, expression (2) reduces to the expression for depolarization field obtained by Kretschmer and Binder.[1]

Fig. 1 schematically illustrates the origin of depolarization fields in thin films with 180°-domain structure and inhomogeneous polarization vector $\mathbf{P}_3(x,z)$. Depolarization field $\mathbf{E}_i^d$ is caused by imperfect screening by the surrounding and inhomogeneous polarization distribution and/or its breaks at interfaces.

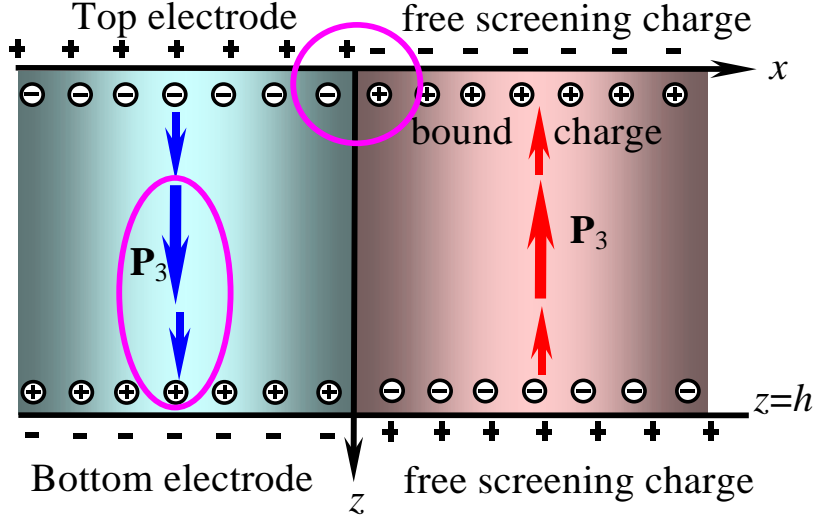

**Fig.1**. (Color online). 180°-domain structure of thin film covered with perfect electrodes. Break of double electric layers at the film-surface junction (marked by a circle) and polarization inhomogeneity (arrows of different length) cause depolarization fields.

Correct phenomenological description of a nanosized system requires the consideration of its surface energy $F_S$. Including the surface energy term, Landau-Ginzburg-Devonshire (LGD) free energy $F = F_S + F_V$ depends on the chosen order parameter – spontaneous polarization component $P_3$.

Within the LGD theory for the II-type incommensurate materials the spatial distribution of the spontaneous polarization component $P_3$ inside the film of thickness $h$ could be found by the minimization of the free energy functional (see e.g. Refs. [36, 37]):

$$F \approx \int_0^h dz \int_{-\infty}^{\infty} dxdy \left( \begin{array}{c} \frac{\alpha'}{2}P_3^2 + \frac{\beta'}{4}P_3^4 + \frac{\gamma}{6}P_3^6 - P_3\left(\frac{E_3^d}{2} + E_3^e\right) \\ + \frac{g_i}{2}\left(\frac{\partial P_3}{\partial x_i}\right)^2 + \frac{w_i}{2}\left(\frac{\partial^2 P_3}{\partial x_i^2}\right)^2 + \frac{v_i}{2} \cdot P_3^2 \left(\frac{\partial P_3}{\partial x_i}\right)^2 \end{array} \right) + \frac{\alpha^S}{2} \int_{-\infty}^{\infty} dxdy \left(P_3^2\big|_{z=0} + P_3^2\big|_{z=h}\right) \quad (2)$$

Coefficient $\alpha(T) = \alpha_T(T - T_C)$ explicitly depends on temperature $T$, $T_C$ is the Curie temperature of a bulk material. Coefficients $\alpha'(T)$ and $\beta'$ may be renormalized due to the electrostriction coupling as shown by Cao and Cross[11] and Zhirnov[10] and misfit strain $u_m^*$ originated from the film and substrate lattice mismatch,[38] namely $\alpha' = \alpha(T) - \dfrac{(Q_{11}+Q_{12})u_m^*}{s_{11}+s_{12}} - 2\dfrac{(Q_{11}^2+Q_{12}^2)s_{11}-2s_{12}Q_{11}Q_{12}}{s_{11}^2-s_{12}^2}\overline{P}_0^2$

and $\beta' = \beta + 2\dfrac{(Q_{11}^2+Q_{12}^2)s_{11}-2s_{12}Q_{11}Q_{12}}{s_{11}^2-s_{12}^2}$, where $s_{ij}$ is the elastic compliance coefficient at constant polarization; $Q_{ij}$ is the electrostriction tensor, $\alpha$ and $\beta$ are stress-free expansion coefficients, the average spontaneous polarization is $\overline{P}_0$ [39]. Note, that the strain $u_m^*$ could depend on the film thickness because of misfit dislocations appearance at some critical thickness $l_d \sim 1/|u_m|$ in accordance with the model proposed by Speck and Pompe.[40]

The coefficients $g_i$ and $\beta$ could be negative, other higher coefficients are positive. Last integral term in Eq.(2) is the surface contribution to the system free energy. Expansion coefficients of the polarization-dependent surface energy may be different for different surface regions. Below we approximate the coordinate dependence by effective value $\alpha^s$ and neglect higher order terms in the surface energy. The depolarization field $E_3^d$ is given by Eq.(2).

In general case the *necessary condition* of the incommensurate phase appearance, $g_i<0$, should be satisfied at least in one spatial direction (i.e. for one value of $i$), while for other directions the homogeneous state would be stable if $g_j>0$. Since most of ferroelectrics with incommensurate phase are uniaxial or biaxial ones, the coefficients $g_i$ are not necessarily equal. In order to simplify our consideration and obtain close-form analytical results, we suppose that $g_3>0$ and either $g_1=g_2<0$ (symmetric biaxial *x,y*-incommensurate case) or $g_1<0 \& g_2>0$ (uniaxial *x*-incommensurate case). In the case one could neglect the higher order derivatives on z and put $w_3=0$ and $v_3=0$ in Eq.(2). Coefficients $w_1$ and $v_1$ should be non-zero positive values for the x-incommensurate modulation existence; or coefficients $w_1=w_2$ and $v_1=v_2$ should be non-zero positive values for the x,y-incommensurate modulation existence. This simplified model allows analytical consideration of the influence of size effects on the incommensurate phase features.

So, under the conditions $g_1=g_2<0$, $w_1=w_2$ and $v_1=v_2$, $g_3>0$ and $w_3=0$ and $v_3=0$, minimization of the free energy (2) results into the relaxation equation for polarization distribution

$$\Gamma\frac{\partial P_3}{\partial t} + \alpha'(T)P_3 + \beta' P_3^3 + \gamma P_3^5 - g_3\frac{\partial^2 P_3}{\partial z^2} - (g_1+v_1 P_3^2)\left(\frac{\partial^2 P_3}{\partial x^2}+\frac{\partial^2 P_3}{\partial y^2}\right)+$$
$$+ w_1\left(\frac{\partial^4 P_3}{\partial x^4}+\frac{\partial^4 P_3}{\partial y^4}\right) - v_1 P_3\left(\left(\frac{\partial P_3}{\partial x}\right)^2+\left(\frac{\partial P_3}{\partial y}\right)^2\right) = E_3^d + E_0^e \exp(i\omega t) \quad (3)$$

Where $\Gamma$ is a positive relaxation coefficient, $\omega$ is the frequency of external field $E_0^e$. For 1D-case one should put $\partial P_3/\partial y = 0$ and consider $P_3(x,z)$ in Eq.(3).

The boundary conditions for polarization acquire the form:

$$\left(\alpha^S P_3 - g_3 \frac{\partial P_3}{\partial x_3}\right)\bigg|_{x_3=0} = 0, \quad \left(\alpha^S P_3 + g_3 \frac{\partial P_3}{\partial x_3}\right)\bigg|_{x_3=h} = 0. \tag{4}$$

Similarly to the case of commensurate ferroelectrics one could introduce extrapolation lengths $\Lambda = g_3/\alpha^S$ that is usually positive. Infinite extrapolation length corresponds to an ideal surface ($\alpha^S \to 0$) and so-called natural boundary conditions, while zero extrapolation length ($\alpha^S \to \infty$) corresponds to $P_3(x_3 = 0) = 0$ at a strongly damaged surface without long-range order. Reported experimental values are $\Lambda = 2 - 50$nm.[41,.42]

### III. Approximate analytical solution of the Euler-Lagrange equations

Then one could find the solution of Eq.(3) linearized for the small polarization modulation $p(\mathbf{k},z)$ in the form of series on the eigen functions $f_n(k,z)$:

$$p(\mathbf{k},z,t) = \sum_n \left(A_n(\mathbf{k})f_n(k,z)\exp\left(-\lambda_n(k)\frac{t}{\Gamma}\right) + E_n(\mathbf{k},\omega)\frac{f_n(k,z)\exp(i\omega t)}{\lambda_n(k) + i\omega\Gamma}\right). \tag{5}$$

Hereinafter $k = |\mathbf{k}|$ and $\mathbf{k} = \{k_x, k_y\}$ for the $x,y$-incommensurate modulation or $\mathbf{k} = \{k_x, 0\}$ for the $x$-incommensurate modulation.

The first term in Eq.(5) is related to the relaxation of initial conditions while the second one is the series expansion external stimulus $\tilde{E}_0^e$ via the eigen functions $f_n(k,z)$. The eigen functions $f_n(k,z)$ and eigen values $\lambda_n(k)$ should be found from the nontrivial solutions of the following problem:

$$\left[\alpha^* - g_3^* \frac{d^2}{dz^2} + g_1^* k^2 + w_1 k^4\right] f_n(k,z) - E_3^d[f_n(k,z)] = \lambda_n(k) f_n(k,z), \tag{6a}$$

$$\left(\alpha^S f_n - g_3 \frac{\partial f_n}{\partial z}\right)\bigg|_{z=0} = 0, \quad \left(\alpha^S f_n + g_3 \frac{\partial f_n}{\partial z}\right)\bigg|_{z=h} = 0. \tag{6b}$$

Where $\alpha^* = \alpha + 3\beta \overline{P}_0^2 + 5\gamma \overline{P}_0^4$, $g_1^* = g_1 + v_1 \overline{P}_0^2$ and $\overline{P}_0$ is the average polarization (for a bulk monodomain sample the spontaneous polarization $P_0^2 = \left(\sqrt{\beta^2 - 4\alpha\gamma} - \beta\right)/2\gamma$). The solution of Eq.(6) was derived as:

$$f_n(k,z) \sim \cosh\left(q_{n1}\left(\frac{z}{h} - \frac{1}{2}\right)\right) - \frac{q_{n2}^2 - h^2 K^2}{q_{n1}^2 - h^2 K^2} \frac{q_{n1} \sinh(q_{n1}/2)}{q_{n2} \sinh(q_{n2}/2)} \cosh\left(q_{n2}\left(\frac{z}{h} - \frac{1}{2}\right)\right), \tag{7}$$

Here $q_{n1,2}$ are expressed via the eigen value $\lambda_n$ as the solutions of the biquadratic equation:

$$\alpha^* + g_1^* k^2 + w_1 k^4 - g_3 \frac{q_n^2}{h^2} + \frac{q_n^2}{\varepsilon_0 \varepsilon_{33} (q_n^2 - h^2 K^2)} = \lambda_n(k). \tag{8}$$

The equation for the eigen spectrum $\lambda_n(k)$ is:

$$\frac{q_{n1} \sinh(q_{n1}/2)}{q_{n1}^2 - h^2 K^2} = \frac{q_{n2} \sinh(q_{n2}/2)}{q_{n2}^2 - h^2 K^2} \left( \frac{\cosh(q_{n1}/2)(\alpha^S/g_3) + (q_{n1}/h)\sinh(q_{n1}/2)}{\cosh(q_{n2}/2)(\alpha^S/g_3) + (q_{n2}/h)\sinh(q_{n2}/2)} \right). \tag{9}$$

Note, that similar equations could be found for "sinh"-eigen functions. Since the smallest (first) eigenvalue should correspond to eigen function of constant sign, we restrict our consideration for the first symmetric "cosh"-eigen functions (7).

The equilibrium dependence of the transverse modulation wave vector $k$ on the temperature $T$ and film thickness $h$ should be found from Eqs.(8)-(9) under the conditions $\lambda_{\min} = 0$.

## IV. Size effect on the phase equilibrium and domain structure temperature evolution
### IV.1. Harmonic approximation

Transcendental Eq.(9) was essentially simplified at the domain structure onset (see Appendix B) so that the approximate expression for the highest and lowest roots was derived in the form:

$$k_\pm^2(h,T) \approx -\frac{g_1^*}{2w_1} \pm \sqrt{\frac{g_1^{*2}}{4w_1^2} - \frac{1}{w_1}\left( \alpha^*(T) + \frac{2\alpha^S g_3}{(\alpha^S \sqrt{g_3 \varepsilon_0 \varepsilon_{33}} + g_3)h + \alpha^S g_3 \varepsilon_0 h^2/4R_d^2} \right)}. \tag{10}$$

Under the typical conditions $R_d \gg 50$ nm and $g_3 \sim 10^{-10}$ J·m$^3$/C$^2$, the term $\alpha^S g_3 \varepsilon_0 h^2/4R_d^2$ can be neglected in the denominator of Eq.(10) without any noticeable loss of precision.

The root $k_-(h,T)$ is always stable only in the incommensurate phase of the bulk material. The root $k_+(h,T)$ can be (meta)stable in thin films even in the temperature range corresponding to the bulk commensurate phase, since domain stripes with definite period correspond to smaller depolarization field in comparison with a monodomain distribution. The direct comparison of the corresponding free energies (2) should be performed in order to determine the film thickness range, where the roots $k_\pm(h,T)$ are stable (and thus monodomain distribution is unstable).

The comparison of the free energies (2) was performed in harmonic approximation. It demonstrated that the root $k_+(h,T)$ can be stable in the wide temperature range starting from the law temperatures (much smaller than $T_C$) the up to the vicinity of the transition temperature into a paraelectric phase. This striking result can be explained in the following way. A monodomain state should be energetically preferable in the commensurate ferroelectric defect-free film placed between perfect conducting electrodes, but only for the case of zero surface energy (coefficient $\alpha^S=0$, $\Lambda\to\infty$). Even under the absence of defects domain stripes originate from imperfect screening (either imperfect electrodes, dielectric gap) and/or the spatial confinement (i.e. surface energy contribution determined by nonzero $\alpha^S$). Zero value of $\alpha^S$ means natural boundary conditions and the absence of size effects, since the polarization is homogeneous along the polar axis and the depolarization field is absent for the case. The non-zero values $\alpha^S$ lead to appearance of polarization inhomogeneity along the polar axis, localized near the surfaces. At the same time, inhomogeneity along the polar axis should induce the depolarization field that affects the periodically modulated domain structure. This effect is a general feature of all ferroics (see e.g. Refs. [43, 3]), it is not related with a bulk incommensurate phase.

The value $\overline{P}_0 \to 0$ in the vicinity of the transition from the paraelectric into modulated ferroelectric phase. So, the transition temperatures into commensurate and incommensurate ferroelectric phases are $T_{CF}(h) = T_C - \dfrac{2\alpha^S g_3}{\alpha_T h\left(\alpha^S \sqrt{g_3\varepsilon_0\varepsilon_{33}} + g_3\right)}$ and $T_{IC}(h) = T_{CF}(h) + \dfrac{g_1^2}{4w_1\alpha_T}$ respectively. At fixed temperature $T$ the transition thicknesses into commensurate and incommensurate phases are $h_{CF}(T) = \dfrac{-2\alpha^S g_3}{\alpha_T(T-T_C)\left(\alpha^S\sqrt{g_3\varepsilon_0\varepsilon_{33}} + g_3\right)}$ and

$h_{IC}(T) = \dfrac{-2\alpha^S g_3}{\left(\alpha_T(T-T_C) - g_1^2/4w_1\right)\left(\alpha^S\sqrt{g_3\varepsilon_0\varepsilon_{33}} + g_3\right)}$ respectively.

As anticipated, Eq.(10) reduces to the well-known bulk solution $k_B^2(T) = -\dfrac{g_1}{2w_1} \pm \sqrt{\dfrac{g_1^2}{4w_1^2} - \dfrac{\alpha_T(T-T_C)}{w_1}}$ with the film thickness $h$ increase. So, for the bulk sample the incommensurate modulation exists in the temperature range $T_C < T < T_{IC}$, where $T_{IC} = T_C + \dfrac{g_1^2}{4w_1\alpha_T}$ and $k_B^2(T_{IC}) = -g_1/2w_1$. Thus, the approximation (10) differs from the bulk solution in renormalization of $\alpha$ by finite size and surface effects.

Phase diagram in coordinates temperature – film thickness with paraelectric (PE), incommensurate (IC) and commensurate (CF) ferroelectric phases, and corresponding domain structure profiles are shown in Fig. 2.

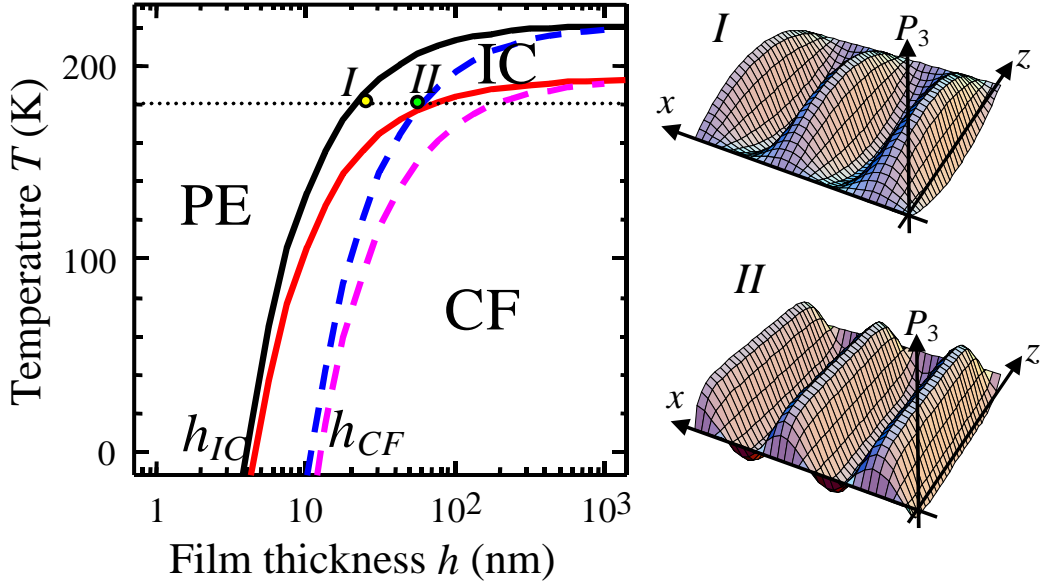

**Fig. 2.** (Color online). Phase diagram in the coordinates temperature-thickness of the film deposed on the matched substrate ($u_m^* = 0$) and placed between perfect conducting electrodes. PE is a paraelectric phase, IC is an incommensurate phase and CF is a commensurate ferroelectric phase. Solid and dashed curves correspond to the surface energy coefficient $\alpha^S = 1$ m$^2$/F and 10 m$^2$/F respectively. Insets schematically show the polarization {x,z}-profiles in the points I and II of the phase diagram. Material parameters of S$_2$P$_2$Se$_6$: $\alpha_T = 1.6 \cdot 10^6$ J·m/(C$^2$·K), $T_C = 193$ K, $\beta = -4.8 \cdot 10^8$ J·m$^5$/C$^4$, $\gamma = 8.5 \cdot 10^{10}$ J·m$^9$/C$^6$, $g_1 = -5.7 \cdot 10^{-10}$ J·m$^3$/C$^2$, $w_1 = 1.8 \cdot 10^{-27}$ J·m$^5$/C$^2$, $v_1 = 1.2 \cdot 10^{-8}$ J·m$^7$/C$^4$, $g_3 = 5 \cdot 10^{-10}$ J·m$^3$/C$^2$ and positive $g_2 \sim g_3$ were taken from Refs.[44, 45], $\varepsilon_{11} = \varepsilon_{33} = 10$ (reference medium is isotropic dielectric).

It is seen from Fig. 2 that the transition temperatures into incommensurate and commensurate phases strongly depend on the film thickness, surface energy and polarization gradient. Thus "soft" incommensurate modulation appeared at thickness $h_{IC}(T)$ and becomes "harder" with the thickness increase in CF (compare insets I and II plotted for thicknesses $h_I < h_{II}$).

Calculated modulation period $q_\pm(h,T) = 2\pi k_\pm^{-1}(h,T)$ is presented in Figs. 3 for different values of the surface energy coefficient $\alpha^S$ (compare curves 1-4). It is seen that in the most cases approximate Eq.(10) (dotted curves) and numerical calculations from Eqs.(7)-(9) (solid curves) give almost the same results.

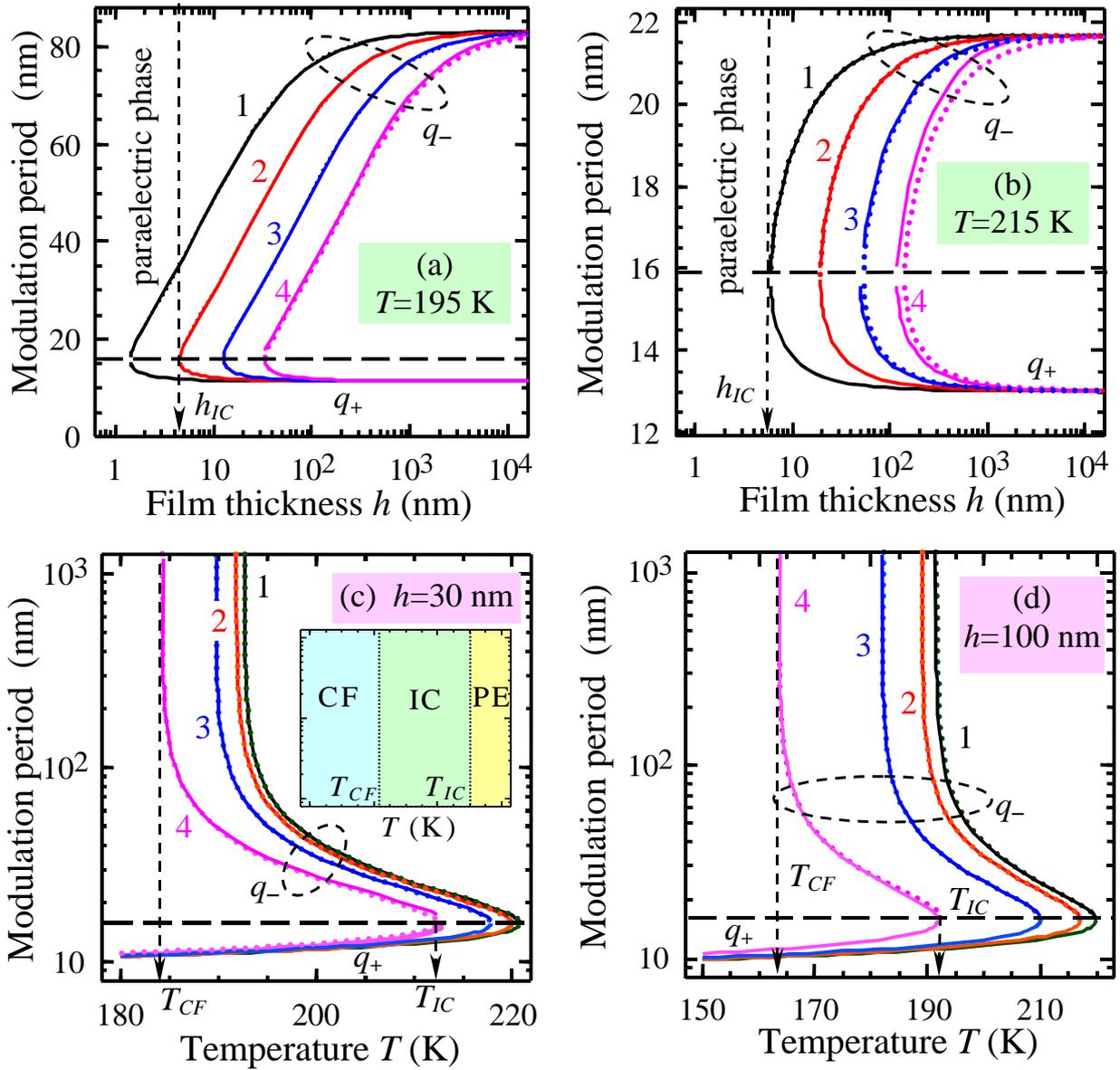

**Fig. 3.** (Color online). Thickness dependences of modulation periods $q_- = 2\pi k_-^{-1}$ (top curves above the dashed horizontal line) and $q_+ = 2\pi k_+^{-1}$ (bottom curves below the dashed horizontal line) for different values of temperature $T = 195$ K (a) and 215 K (b). Temperature dependences of $q_- = 2\pi k_-^{-1}$ (top curves) and $q_+ = 2\pi k_+^{-1}$ (bottom curves) for different values of the film thickness $h = 30$ nm (c) and 100 nm (d). Curves 1, 2, 3, 4 correspond to the surface energy coefficient $\alpha^S = 0.03$, 0.1, 0.3, 1 m$^2$/F respectively. Solid and dotted curves represent exact numerical calculations from Eqs.(7)-(9) and approximate analytical dependences (10) respectively. Material parameters are the same as in Fig. 2 and $R_d \sim 500$ nm.

Summarizing results obtained in this section, we would like to underline that transition temperatures $T_{CF}$, $T_{IC}$ and the maximal period $q_- = 2\pi k_-^{-1}$ of the incommensurate domain structure strongly depend on the film thickness and surface energy, while the minimal period $q_+ = 2\pi k_+^{-1}$ weakly depend on the film thickness and surface energy. The correlation effects, which strength is in turn determined from the value of gradient coefficient $g_1^*$, determine the scale of both periods. The dependence of all polar properties on the Debye screening radius $R_d$ is rather weak under the typical conditions $R_d \gg 50$ nm.

**IV.2. Phase-field modeling**

In order to check the validity of the analytical calculations of performed in harmonic approximation, we study the problem numerically by the phase-field modeling method. In the phase-field approach [46, 47, 48], we use the spatial distribution of the spontaneous polarization to describe the domain structure. The distribution of electric field is obtained by solving the electrostatic equations supplemented by the boundary conditions at the top and bottom electrodes. All-important energetic contributions (including the electrostatic energy and surface energy) are incorporated into the total LGD free energy functional $F(P_1, P_2, P_3, u_{ij})$. The temporal evolution of the polarization vector field, and thus the domain structure, is then described by the time-dependent LGD equations $\frac{\partial P_i}{\partial t} = -\Gamma \frac{\delta F}{\delta P_i}$, where $\Gamma$ is the kinetic coefficient related to the domain-wall mobility. For a given initial distributions, numerical solution of time-dependent LGD equations yields the temporal and spatial evolution of the polarization. We use periodic boundary conditions along both the x and y directions.

Approximate analytical results are compared with numerical phase-field calculations of 2D {x,y}-modulated domain structures in Figs. 4 for a thin film with dimensions 100 x 100 x 40 nm at different temperatures. A positive surface energy coefficient $\alpha^S$ was employed. It is seen from the plots (b)-(e) that domain structure originated at low temperatures. So, the monodomain state appeared unstable starting from much lower temperatures than $T_{CF}$. This supports the assumption made in the section V.1 that domain stripes in ferroelectric phase possibly originate from finite surface energy value determined by nonzero $\alpha^S$. It should be emphasized that periodic boundary conditions along both the x and y directions should affect the periodicity of the incommensurate structures.

We also performed 1D-phase field simulations to calculate the incommensurate /x/-modulated structures in thin_ plates with sizes $h_x = 250$ nm, $h_y = 2$ nm, $h_z = 40$ nm. We calculated the polarization distribution at different temperatures [Fig.5]. From Fig. 5 it can be seen that the $P_3$ distribution across the film depth $z$ looks like a dome at fixed $x$ position. $P_3$ is maximal in the middle of the film, and its module decreases from middle to edges. Transversal $x$-distribution of $P_3$ is periodic and it looks like sine wave with a period about 17.6 nm (18 grids) [Figs. 5c-h].

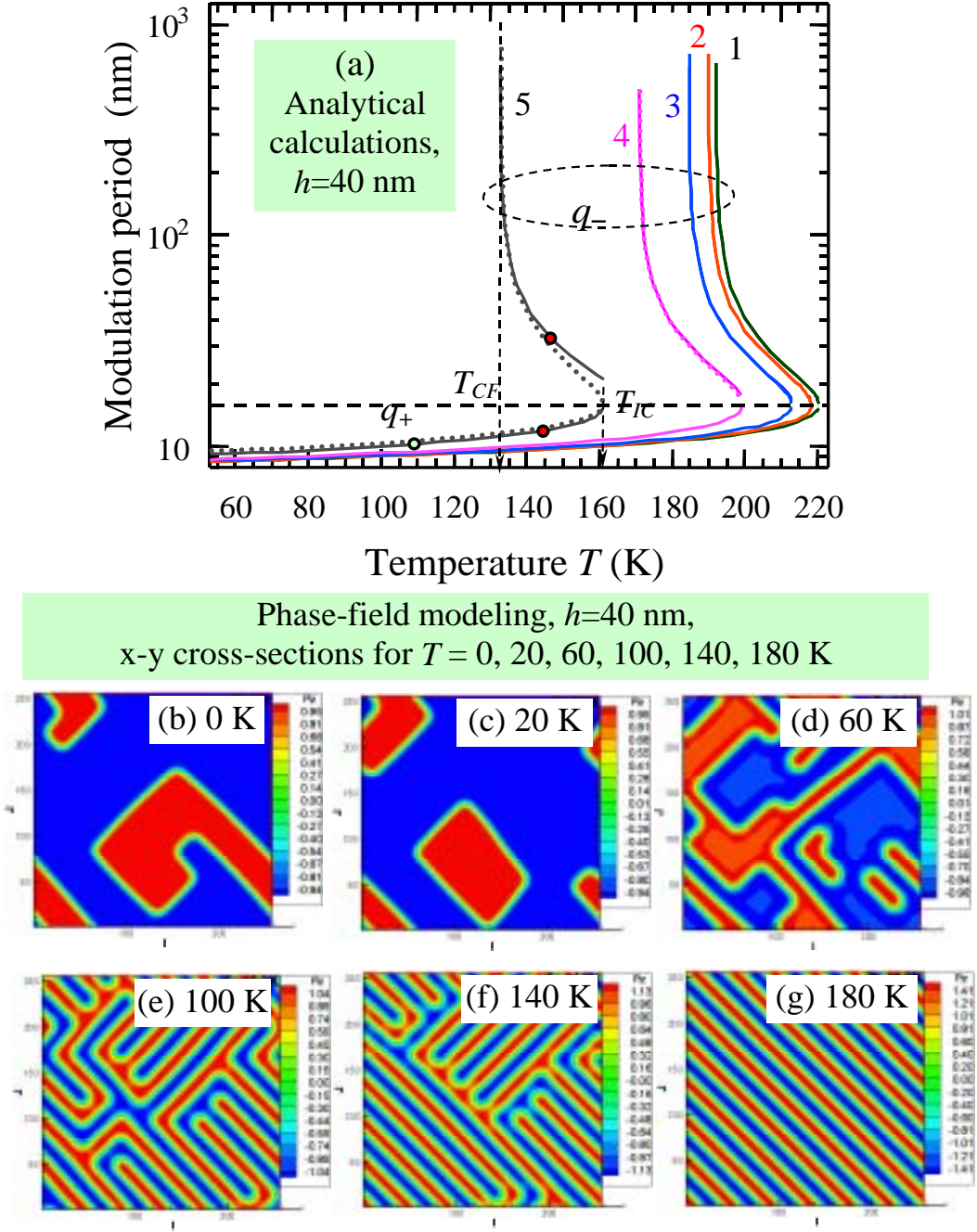

**Fig. 4.** (Color online). (a) Temperature dependence of the modulation period $q_-$ (top curves above the dashed horizontal line) and $q_+$ (bottom curves below the dashed horizontal line) calculated analytically for the film thickness $h = 40$ nm. Curves 1, 2, 3, 4, 5 correspond to the surface energy coefficient $\alpha^S = 0.03, 0.1, 0.3, 1, 10$ m$^2$/F respectively. Solid and dotted curves represent numerical calculations from Eqs.(7)-(9) and approximate analytical dependences (10) respectively. (b-e) Temperature evolution of the {x,y}-modulated domain structure calculated by the phase-field modeling for the film with sizes 100 x 100 x 40 nm, $\alpha^S = 10$ m$^2$/F and temperatures $T = 0, 20, 60, 100, 140$ and $180$ K. Other parameters are the same as in Fig. 2, but $g_2 = g_1 = -5.7 \cdot 10^{-10}$ J·m$^3$/C$^2$ and $R_d \sim 500$ nm.

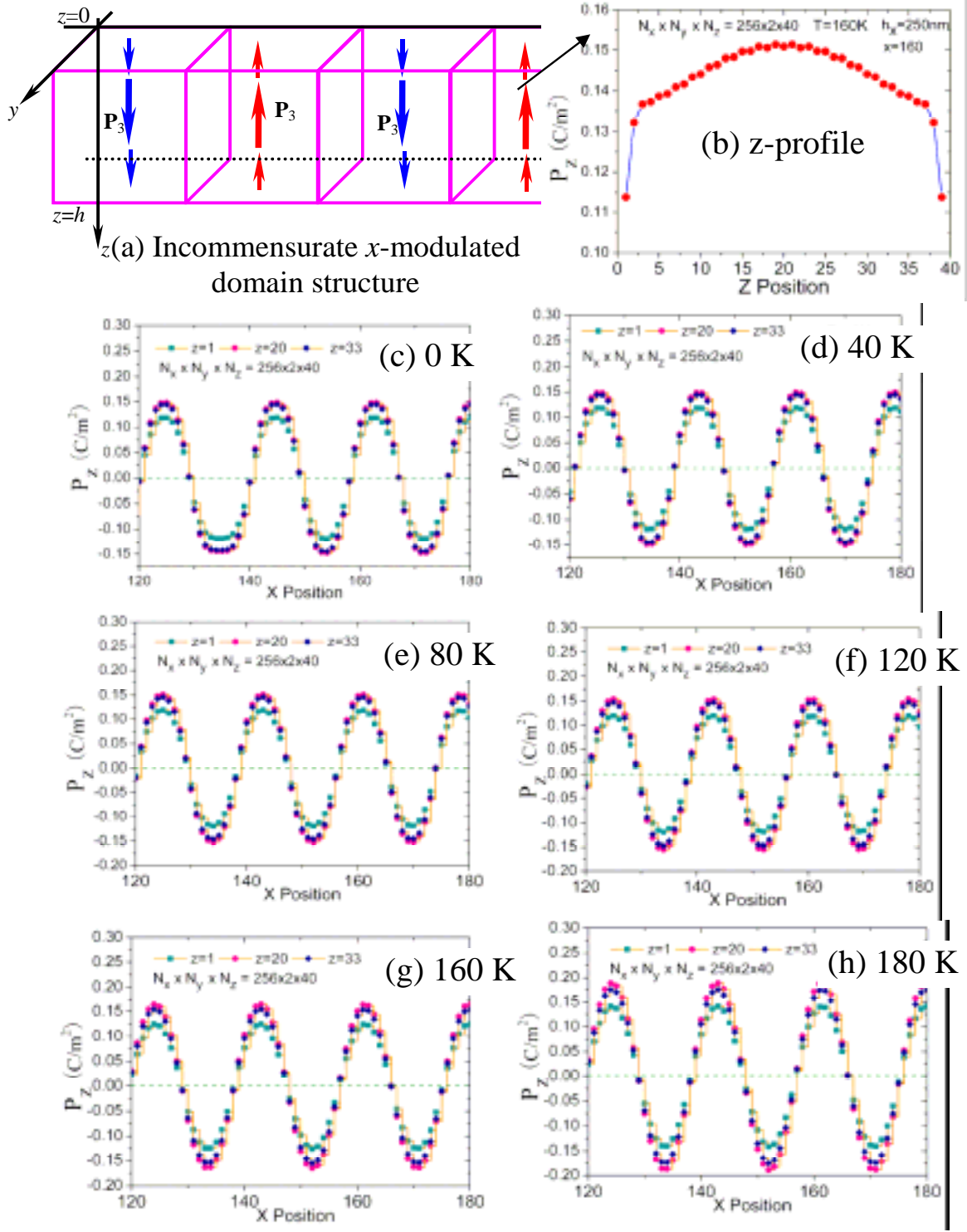

**Fig. 5.** (Color online). (a) Sketch of the 1D *x*-modulated domain structure. (b) Phase field 1D-simulation of the polarization component $P_3$ variation for different z positions at $T = 160$ K and $x=160$. (c-h) Polarization component $P_3$ morphologies in thin plates with sizes $h_x = 250$ nm, $h_y = 2$ nm and $h_z = 40$ nm at temperatures $T = 0, 40, 80, 120, 160$ and $180$ K. The pink solid circle, navy rhombus and dark cyan square represent $P_3$ at z=20, z=33 and z=1 respectively. Other parameters are the same as in Fig. 4, but $g_1 = -5.7 \cdot 10^{-10}$ J·m$^3$/C$^2$ and $g_2 > 0$.

## V. Summary

We proposed the theoretical description of finite size, depolarization effects, surface and correlation energy influence on the phase diagram of thin ferroelectric films with II-type incommensurate phases and semiconductor properties.

Within the framework of Landau-Ginzburg-Devonshire theory we performed analytical calculations and phase-field modeling of the temperature evolution and thickness dependence of the period of incommensurate $180^o$-domains appeared in thin films covered with perfect electrodes. Despite numerous efforts, the problem has not been solved previously.

It was shown analytically that the transition temperature between paraelectric, incommensurate and commensurate ferroelectric phases (as well as the period of incommensurate domain structures) strongly depend on the film thickness, surface energy and gradient coefficients. At the same time their dependences on Debye screening radius $R_d$ are rather weak for the typical values $R_d \gg 50$ nm.

Unexpectedly, both the analytical theory and phase-field modeling results demonstrate that the incommensurate modulation weakly dependent on temperature can be stable in thin films in the wide temperature range starting from the low temperatures (much smaller than the bulk Curie temperature) up to the temperature of paraelectric phase transition.

These domain stripes possibly originate even at low temperatures from the spatial confinement and finite surface energy contribution. Non-zero values of the surface energy lead to appearance of polarization inhomogeneity along the polar axis, localized near the surfaces. This effect could be the common feature of various confined ferroelectrics and ferromagnetics. Thus we expect that the long-range order parameter (e.g. spontaneous polarization or magnetization) subjected to either spatial confinement or imperfect screening could reveal incommensurate modulation in nanosized ferroics. The result can be important for novel applications of the nanosized materials in nanoelectronics and memory devices.

**Acknowledgements**

Research sponsored by Ministry of Science and Education of Ukraine and National Science Foundation (Materials World Network, DMR-0908718).

**Appendix A. Calculations of depolarization field and linearized solution**

Variation of the Helmholtz energy (2) on polarization leads to the following equations of state $\frac{\partial F}{\partial P_3} - \frac{\partial}{\partial x_i}\frac{\partial F}{\partial(\partial P_3/\partial x_i)} + \frac{\partial^2}{\partial x_i^2}\frac{\partial F}{\partial(\partial^2 P_3/\partial x_i^2)} = 0$, namely:

$$\Gamma\frac{\partial P_3}{\partial t} + \alpha'(T)P_3 + \beta' P_3^3 + \gamma P_3^5 - g_3\frac{\partial^2 P_3}{\partial z^2} - (g_1 + v_1 P_3^2)\left(\frac{\partial^2 P_3}{\partial x^2} + \frac{\partial^2 P_3}{\partial y^2}\right) + $$
$$+ w_1\left(\frac{\partial^4 P_3}{\partial x^4} + \frac{\partial^4 P_3}{\partial y^4}\right) - v_1 P_3\left(\left(\frac{\partial P_3}{\partial x}\right)^2 + \left(\frac{\partial P_3}{\partial y}\right)^2\right) = E_3^d + E_3^e \quad (A.1a)$$

Where $\Gamma$ is positive relaxation coefficient. The boundary conditions for polarization in the form

$$\left(\alpha_0^S P_3 - g_3\frac{\partial P_3}{\partial x_3}\right)\bigg|_{x_3=0} = 0, \quad \left(\alpha_h^S P_3 + g_3\frac{\partial P_3}{\partial x_3}\right)\bigg|_{x_3=h} = 0. \quad (A.1b)$$

Similarly to the case of commensurate ferroelectrics one could introduce extrapolation lengths $\Lambda_{1,2} = g_3/\alpha_{0,h}^S$ which are usually regarded positive. Infinite extrapolation length corresponds to ideal surface ($\alpha_{0,h}^S \to 0$) and so-called natural boundary conditions, while zero extrapolation length ($\alpha_{0,h}^S \to \infty$) corresponds to $P_3(x_3 = 0) = 0$ at strongly damaged surface without long-range order. Below we mainly consider the case of equal values $\alpha_0^S = \alpha_h^S$ for the sake of simplicity.

The Maxwell's equations for the inner electric field $\mathbf{E}_i = -\nabla\varphi_i(\mathbf{r})$, expressed via electrostatic potential $\varphi_i(\mathbf{r})$ and polarization $\mathbf{P}(\mathbf{r})$ with supplementary boundary conditions have the form:

$$\begin{cases} \text{div}(\mathbf{P}(\mathbf{r}) - \varepsilon_0\nabla\varphi_i(\mathbf{r})) = \rho(\mathbf{r}), \quad z \geq 0, \\ \varphi_i(x,y,0) = \varphi_e(x,y,0), \quad \varphi_i(x,y,h) = 0. \end{cases} \quad (A.2)$$

Potential $\varphi_e(\mathbf{r})$ is created by the electrode located at the sample surface. Electrostatic potential $\varphi_i(\mathbf{r})$ includes bond charges (electric depolarization field) and free charges $\rho(\mathbf{r})$; $\varepsilon_0$ is the dielectric constant, $h$ is the film thickness. The perfect screening of depolarization field outside the sample is realized by the ambient charges or electrodes, i.e. there is no dielectric gap.

In Debye approximation we can rewrite the problem (5) for electrostatic potential as:

$$\begin{cases} \varepsilon_{33}\frac{\partial^2\varphi_i}{\partial z^2} + \varepsilon_{11}\left(\frac{\partial^2\varphi_i}{\partial x^2} + \frac{\partial^2\varphi_i}{\partial y^2}\right) - \frac{\varphi_i}{R_d^2} = \frac{1}{\varepsilon_0}\frac{\partial P_3}{\partial z}, \\ \varphi_i(x,y,0) = \varphi_e(x,y,t), \quad \varphi_i(x,y,h) = 0 \end{cases} \quad (A.3)$$

$R_d$ is the Debye screening radius, background permittivity, $\varepsilon_{33}$ is regarded much smaller than ferroelectric contribution to permittivity, $\varepsilon_{33}^f$. Corresponding Fourier representation on transverse coordinates $\{x,y\}$ for electrostatic potential $\tilde{\varphi}_i(\mathbf{k},z)$ and electric field normal component $\tilde{E}_3 = -\partial \tilde{\varphi}_i/\partial z$ have the form:

$$\tilde{\varphi}_i(\mathbf{k},z,t) = \tilde{\varphi}_e(\mathbf{k},t)\frac{\sinh(K(h-z))}{\sinh(Kh)} + \left(\begin{array}{l} \int_0^z dz' \tilde{P}_3(\mathbf{k},z')\frac{\cosh(Kz')\sinh(K(h-z))}{\varepsilon_0 \varepsilon_{33} \cdot \sinh(Kh)} - \\ \int_z^h dz' \tilde{P}_3(\mathbf{k},z')\frac{\sinh(Kz)\cosh(K(h-z'))}{\varepsilon_0 \varepsilon_{33} \cdot \sinh(Kh)} \end{array}\right), \quad (A.4)$$

$$\tilde{E}_3(\mathbf{k},z,t) = \tilde{E}_3^e(\mathbf{k},z,t) + \tilde{E}_3^d(\mathbf{k},z), \quad \tilde{E}_3^e(\mathbf{k},z,t) = \tilde{\varphi}_e(\mathbf{k},t)\frac{\cosh(K(h-z))}{\sinh(Kh)}K(k),$$

$$\tilde{E}_3^d[\tilde{P}_3(\mathbf{k},z)] = \left(\begin{array}{l} -\frac{1}{\varepsilon_0 \varepsilon_{33}}\tilde{P}_3(\mathbf{k},z) + \int_0^z dz' \tilde{P}_3(\mathbf{k},z')\cosh(Kz')\frac{\cosh(K(h-z))}{\varepsilon_0 \varepsilon_{33} \cdot \sinh(Kh)}K + \\ \int_z^h dz' \tilde{P}_3(\mathbf{k},z')\cosh(K(h-z'))\frac{\cosh(Kz)K(k)}{\varepsilon_0 \varepsilon_{33} \cdot \sinh(Kh)} \end{array}\right) \quad (A.5)$$

Here vector $\mathbf{k} = \{k_1, k_2\}$, its absolute value $k = \sqrt{k_1^2 + k_2^2}$ and so $K(k) = \sqrt{(\varepsilon_{11}k^2 + R_d^{-2})/\varepsilon_{33}}$. $\tilde{\varphi}_e(\mathbf{k})$ is the external electric field potential at the sample surface. Ever under the perfect external screening or short-circuit condition $\tilde{\varphi}_e(\mathbf{k}) = 0$, the field (8) is nonzero; it is produced by inhomogeneous polarization distribution, i.e. it is typical depolarization field.

When the dependence of depolarization field on polarization is known, one can find the domain stripes period by the following way.

Fourier representation on transverse coordinates $\{x,y\}$ of Eqs. (A.1) linearized for the small polarization modulation $p(\mathbf{k},z)$ has the form

$$\left[\Gamma\frac{\partial}{\partial t} + \alpha^* - g_3\frac{d^2}{dz^2} + g_1^* k^2 + w_1 k^4\right] p(\mathbf{k},z) = E_3^d[p(\mathbf{k},z)] + \tilde{E}_3^e(\mathbf{k},z)\exp(i\omega t) \quad (A.6)$$

Where $\alpha^* = \alpha + 3\beta \overline{P}_0^2 + 5\gamma \overline{P}_0^4$ and $g_1^* = g_1 + v_1 \overline{P}_0^2$. Boundary conditions to Eq.(A.7) are

$$\left(\alpha^S p - g_3 \frac{\partial p}{\partial x_3}\right)\bigg|_{x_3=0} = 0, \quad \left(\alpha^S p + g_3 \frac{\partial p}{\partial x_3}\right)\bigg|_{x_3=h} = 0. \quad (A.8)$$

One could find the solution of Eq.(A.8) in the form of expansion on the eigen functions $f_n(k,z)$:

$$p(\mathbf{k},z) = \sum_n \left(A_n(\mathbf{k})f_n(k,z)\exp\left(-\lambda_n(k)\frac{t}{\Gamma}\right) + E_n(\mathbf{k},\omega)\frac{f_n(k,z)\exp(i\omega t)}{\lambda_n(k) + i\omega \Gamma}\right). \quad (A.9)$$

Here the first term is related to the relaxation of initial conditions while the second one is the series expansion external stimulus $\tilde{E}_3^e$ via the eigen functions $f_n(k,z)$. The eigen functions $f_n(k,z)$ and positive eigenvalues $\lambda_n(k)$ should be found from the nontrivial solutions of the following problem

$$\left[\alpha^* - g_3^* \frac{d^2}{dz^2} + g_1^* k^2 + w_1 k^4\right] f_n(k,z) - E_3^d[f_n(k,z)] = \lambda_n(k) f_n(k,z) \tag{A.10}$$

$$\left(\alpha^S f_n - g_3 \frac{\partial f_n}{\partial z}\right)\bigg|_{z=0} = 0, \quad \left(\alpha^S f_n + g_3 \frac{\partial f_n}{\partial z}\right)\bigg|_{z=h} = 0. \tag{A.11}$$

As it follows from Eq.(A.11), the following relation is valid $\varepsilon_0 \varepsilon_{33} \partial^2 E_3^d[f(z)]/\partial z^2 - \varepsilon_0(\varepsilon_{11} k^2 + R_d^{-2}) E_3^d[f(z)] = -\partial^2 f(z)/\partial z^2$, so it could be rewritten as

$$\left(\varepsilon_0 \varepsilon_{33} \frac{d^2}{dz^2} - \varepsilon_0\left(\varepsilon_{11} k^2 + \frac{1}{R_d^2}\right)\right)\left[\alpha^* - \lambda_n(k) - g_3 \frac{d^2}{dz^2} + g_1^* k^2 + w_1 k^4\right] f_n(z) + \frac{d^2 f_n(z)}{dz^2} = 0 \tag{A.12}$$

The solution of Eq.(A.12) is obvious

$$f_n(z) = \sum_{i=1}^{2}\left(a_{ni} \sinh\left(q_{ni}\left(\frac{z}{h} - \frac{1}{2}\right)\right) + b_{ni} \cosh\left(q_{ni}\left(\frac{z}{h} - \frac{1}{2}\right)\right)\right), \tag{A.13}$$

Here $q_{ni}$ are the solutions of the following characteristic equation

$$\alpha^* + g_1^* k^2 + w_1 k^4 - g_3 \frac{q_n^2}{h^2} + \frac{q_n^2}{\varepsilon_0 \varepsilon_{33}(q_n^2 - h^2 K^2)} = \lambda_n(k). \tag{A.14}$$

Here $K = \sqrt{(\varepsilon_{11} k^2 + R_d^{-2})/\varepsilon_{33}}$.

Using evident form (2) of depolarization field, one could find the following identities

$$\tilde{E}_3^d\left[\cosh\left(q\left(\frac{z}{h} - \frac{1}{2}\right)\right)\right] = -\frac{q^2 \cosh(q(1/2 - z/h))}{\varepsilon_0(\varepsilon_{33} q^2 - h^2(\varepsilon_{11} k^2 + R_d^{-2}))} + \frac{q h K \sinh(q/2)\cosh(K(z - h/2))}{\varepsilon_0 \varepsilon_{33}(q^2 - h^2 K^2)\sinh(K h/2)}, \tag{A.15a}$$

$$\tilde{E}_3^d\left[\sinh\left(q\left(\frac{z}{h} - \frac{1}{2}\right)\right)\right] = -\frac{q^2 \sinh(q(1/2 - z/h))}{\varepsilon_0(\varepsilon_{33} q^2 - h^2(\varepsilon_{11} k^2 + R_d^{-2}))} + \frac{q h K \cosh(q/2)\sinh(K(z - h/2))}{\varepsilon_0 \varepsilon_{33}(q^2 - h^2 K^2)\cosh(K h/2)}. \tag{A.15b}$$

So, the substitution of the solution (A.13) into Eq. (A.10) with respect to relations (A.14) and (A.15) gives the condition of zero "imbalance":

$$\frac{\sinh(K(z - h/2))}{\varepsilon_0 \varepsilon_{33} \cosh(K h/2)} \sum_{i=1}^{2} \frac{a_{ni} q_{ni} h K}{q_{ni}^2 - h^2 K^2} \cosh\left(\frac{q_{ni}}{2}\right) + \frac{\cosh(K(z - h/2))}{\varepsilon_0 \varepsilon_{33} \sinh(K h/2)} \sum_{i=1}^{2} \frac{b_{ni} q_{ni} h K}{q_{ni}^2 - h^2 K^2} \sinh\left(\frac{q_{ni}}{2}\right) = 0 \tag{A.16}$$

Since hyperbolic sin and cos-functions are linearly independent, validity of Eq (A.16) at any coordinate $z$ is possible under conditions:

$$\sum_{i=1}^{2} a_{ni} \frac{q_{ni} \cosh(q_{ni}/2)}{q_{ni}^2 - h^2 K^2} = 0, \tag{A.17a}$$

$$\sum_{i=1}^{2} b_{ni} \frac{q_{ni} \sinh(q_{ni}/2)}{q_{ni}^2 - h^2 K^2} = 0. \tag{A.17b}$$

Boundary conditions (A.11) give the following system of equations:

$$\sum_{i=1}^{2} a_{ni}\left(\sinh\left(\frac{q_{ni}}{2}\right)\alpha^S + g_3 \frac{q_{ni}}{h}\cosh\left(\frac{q_{ni}}{2}\right)\right) = 0, \quad (A.18a)$$

$$\sum_{i=1}^{2} b_{ni}\left(\cosh\left(\frac{q_{ni}}{2}\right)\alpha^S + g_3 \frac{q_{ni}}{h}\sinh\left(\frac{q_{ni}}{2}\right)\right) = 0. \quad (A.18b)$$

Thus Eqs. (A.17a), (A.18a) and (A.17b), (A.18b) give two systems of equations for the determination of $a_{ni}$ and $b_{ni}$ respectively. Since all these equations are homogeneous, the nontrivial solutions are possible only under conditions of zero determinant. Considering cosh-distribution (from the system of equations for coefficients $b_{ni}$), one could find the following relation:

$$\frac{q_{n1}\sinh(q_{n1}/2)}{q_{n1}^2 - h^2 K^2}\left(\cosh\left(\frac{q_{n2}}{2}\right)\alpha^S + g_3 \frac{q_{n2}}{h}\sinh\left(\frac{q_{n2}}{2}\right)\right) =$$
$$= \frac{q_{n2}\sinh(q_{n2}/2)}{q_{n2}^2 - h^2 K^2}\left(\cosh\left(\frac{q_{n1}}{2}\right)\alpha^S + g_3 \frac{q_{n1}}{h}\sinh\left(\frac{q_{n1}}{2}\right)\right) \quad (A.19)$$

Introducing the designations $s_{ni}^2 = q_{ni}^2/h^2$, one could rewrite Eqs. (A.14), (A.19) in the compact form as:

$$s_{n1,2}^2 = \frac{1}{2}\left(K^2 + \frac{Q_n \varepsilon_0 \varepsilon_{33} + 1}{g_3 \varepsilon_0 \varepsilon_{33}} \pm \sqrt{\left(K^2 + \frac{Q_n \varepsilon_0 \varepsilon_{33} + 1}{g_3 \varepsilon_0 \varepsilon_{33}}\right)^2 - \frac{4K^2 Q_n}{g_3}}\right) \quad (A.20a)$$

$$\frac{s_{n1}\sinh(s_{n1}h/2)}{s_{n1}^2 - K^2}\left(\cosh\left(\frac{s_{n2}h}{2}\right)\alpha^S + g_3 s_{n2}\sinh\left(\frac{s_{n2}h}{2}\right)\right) =$$
$$= \frac{s_{n2}\sinh(s_{n2}h/2)}{s_{n2}^2 - K^2}\left(\cosh\left(\frac{s_{n1}h}{2}\right)\alpha^S + g_3 s_{n1}\sinh\left(\frac{s_{n1}h}{2}\right)\right) \quad (A.20b)$$

Here $Q_n(k) = \alpha^* + g_1^* k^2 + w_1 k^4 - \lambda_n$. It is seen that one of the roots, $s_{n2}$, corresponding to sign plus, is always real and positive number, while the root $s_{n1}$, corresponding to sign minus, could either real or imaginary, depending on $Q_n$ sign.

Similar equations could be found for sinh-distribution (from the system of equations for coefficients $a_{ni}$), however, since the smallest (first) eigenvalue should correspond to eigenfunction of constant sign, we restrict our consideration for first symmetric eigenfunction.

**Appendix B. Derivation of Eq.(10)**

At small $K$ values the following limits of Eq.(A.20a) are valid

$$s_{n1}^2 \to \frac{K^2 Q_n \varepsilon_0 \varepsilon_{33}}{Q_n \varepsilon_0 \varepsilon_{33} + 1}, \quad s_{n2}^2 \to \frac{Q_n \varepsilon_0 \varepsilon_{33} + 1}{g_3 \varepsilon_0 \varepsilon_{33}} + \frac{K^2}{Q_n \varepsilon_0 \varepsilon_{33} + 1}. \quad (B.1)$$

At high $s_2 h$ and small $s_1 h$ values (which is valid for not very high wave vectors and intermediate values of film thickness) determinant (A.20b) is

$$\left(\frac{s_{n1}^2 h/2}{s_{n1}^2 - K^2}\left(\alpha^S + g_3 s_{n2}\right) - \frac{s_{n2}}{s_{n2}^2 - K^2}\left(\alpha^S + g_3 s_{n1}^2 \frac{h}{2}\right)\right)\exp\left(\frac{s_{n2}h}{2}\right) = 0. \quad (B.2)$$

This expression could be further simplified by expanding in series on $Q_n$ and reduced to

$$Q_n + 2\alpha^S g_3 \left(h\left(g_3 + \frac{\alpha^S \sqrt{g_3 \varepsilon_0 \varepsilon_{33}}}{\left(1 + g_3 \varepsilon_0 R_d^{-2}\right)^{3/2}}\right) + \frac{\alpha^S g_3 \varepsilon_0}{4\left(1 + g_3 \varepsilon_0 R_d^{-2}\right)}\frac{h^2}{R_d^2}\right)^{-1} = 0 \quad (B.3)$$

Under the condition $\lambda_n = 0$ and inequality $g_3 \varepsilon_0 R_d^{-2} \ll 1$ (typically valid with high accuracy at $R_d > 5$ nm and $g_3 \sim 10^{-10}$ J·m$^3$/C$^2$, $\varepsilon_0 = 8.85 \cdot 10^{-12}$ F/m) the Eq.(B.3) leads to the following equation for the wave vector $k$:

$$\alpha^* + g_1^* k^2 + w_1 k^4 + \frac{2\alpha^S g_3}{\left(\alpha^S \sqrt{g_3 \varepsilon_0 \varepsilon_{33}} + g_3\right)h + \alpha^S g_3 \varepsilon_0 h^2/4R_d^2} = 0 \quad (B.4)$$

that leads to the following expression

$$k_h^2(h,T) = -\frac{g_1^*}{2w_1} \pm \sqrt{\frac{g_1^{*2}}{4w_1^2} - \frac{1}{w_1}\left(\alpha(T) + \frac{2\alpha^S g_3}{\left(\alpha^S \sqrt{g_3 \varepsilon_0 \varepsilon_{33}} + g_3\right)h + \alpha^S g_3 \varepsilon_0 h^2/4R_d^2}\right)} \quad (B.5)$$

[35] The dependence of $\varepsilon_{11}$ on $E_3$ is absent for uniaxial ferroelectrics. It may be essential for perovskites with high coupling constant.